\documentclass[10pt]{article}

\usepackage{amsmath} \usepackage{amssymb} \usepackage{graphics}

\usepackage{tikz} \usetikzlibrary{calc} \usetikzlibrary{arrows}

\newenvironment{Diagram}[3][1]{\begin{tikzpicture} [scale=#1*0.25, baseline=-5-#3cm] \path (0,0)+(#2*4/#1,0) coordinate (XYshift);}{\end{tikzpicture}}

\newcommand{\awire}[2]{\draw[->] (#1) -- ($0.5*(#1)+ 0.5*(#2)$) coordinate (fontlocation); \draw ($0.5*(#1)+ 0.5*(#2)$) -- (#2);}

\newcommand{\point}[2]{\path ($(#2)+(XYshift)$) coordinate (#1);}

\newcommand{\beamsplitter}[2][0]{\draw[very thick] ($(#2)+(#1:-1.5)$) -- ($(#2)+(#1:1.5)$); \path (#2) coordinate (fontlocation);}

\newcommand{\mirror}[2][1]{\draw[very thick] ($(#2)+(#1:-1.2)$) -- ($(#2)+(#1:1.2)$); \path (#2) coordinate (fontlocation);}

\newcommand{\source}[2][0]{\draw[thick] ($(#2)+(#1:1.5cm)$) circle (1.5 cm) ; \path ($(#2)+(#1:1.5cm)$) coordinate (fontlocation);}

\newcommand{\detector}[2][0]{\draw[thick] ($(#2)+(#1:2cm)$) arc (#1:#1+180:2cm) -- cycle; \path (#2) coordinate (fontlocation); }

\newcommand{\phaseshifter}[2][0]{\draw[<->] (#2) -- ++(#1:2cm); \draw[thick] ($(#2) +(#1:2.4cm) + (#1+90:3mm)$) -- ++(#1: 2cm) -- ++(#1-90:6mm) -- ++(#1: -2cm) -- cycle ; \path (#2) ++(#1:
2.9cm) coordinate (fontlocation);}

\newcommand{\symb}[2][0,0]{\path ($(fontlocation) + (#1)$) node {\ensuremath{#2}};}

\title{\textbf{Are quantum states real?}} \author{Lucien Hardy\\ \textit{Perimeter Institute,}\\ \textit{31 Caroline Street North,}\\ \textit{Waterloo, Ontario N2L 2Y5, Canada}}

\begin{document}

\maketitle

\begin{abstract} In this paper we consider theories in which reality is described by some underlying variables, $\lambda$.  Each
value these variables can take represents an \emph{ontic state} (a particular state of reality). The preparation of a quantum state corresponds to a distribution over the ontic states,
$\lambda$.   If we make three basic assumptions, we can show that the distributions over ontic states corresponding to distinct pure states are non-overlapping.  This means that we can deduce the quantum state from a knowledge of the ontic state. Hence, if these assumptions are correct, we can claim that the quantum state is a real thing (it is written into the underlying variables that describe reality).  The key assumption we use in this proof is \emph{ontic indifference} - that quantum transformations that do not affect a given pure quantum state can be implemented in such a way that they do not affect the ontic states in the support of that state.  In fact this assumption is violated in the Spekkens toy model (which captures many aspects of quantum theory and in which different pure states of the model have overlapping distributions over ontic states).   This paper proves that ontic indifference must be violated in any model reproducing quantum theory in which the quantum state is not a real thing.  The argument presented in this paper is different from that given in a recent paper by Pusey, Barrett, and Rudolph.  It uses a different key assumption and it pertains to a single copy of the system in question.
\end{abstract}

\section{Introduction} \label{introduction}

Ever since the Schroedinger wrote down his wave equation, people have wondered what kind of thing the wave function is.  Perhaps the most basic question one can ask is whether it is a real
thing.  That is, is there something in the underlying reality of the world that corresponds to the wavefunction?

We start by assuming there is still some underlying reality, described by some variables, $\lambda$.  Sometimes these variables are called \emph{hidden variables}.  This is a little
inappropriate though since not all the variables have to be \lq\lq hidden" \cite{bell2004speakable}.  Each $\lambda$ describes a possible \emph{ontic state} - a possible state of the
underlying reality.  When a quantum state is prepared we imagine that it corresponds to a probability distribution over ontic states.  A different quantum state would, in general, correspond
to a different distribution over ontic states.

Spekkens has pioneered the idea that we might be able to provide an epistemic interpretation of the quantum state.   He has constructed a toy theory \cite{spekkens2007evidence} (see also
\cite{hardy1999disentangling}, \cite{coecke2011toy, coecke2011phase}, \cite{bartlett2011reconstruction}) which is capable of reproducing a substantial fraction of the predictions of quantum
theory.  This toy theory has the property that, for different pure states in the theory, the corresponding distribution over ontic states overlap.  Thus, there exist ontic states that are
consistent with two or more pure states in this toy theory.   The toy theory does not reproduce quantum theory however.  A natural question to ask is whether there exists a model having this
property that can reproduce the predictions of quantum theory.

What if we can prove that pure states, in quantum theory, must correspond to non-overlapping distributions over ontic states?  This would imply that, no matter how complicated the variables,
$\lambda$,  representing the ontic state are, we would be able to deduce from $\lambda$ what the pure quantum state is (since each $\lambda$ would only appear in the distribution of one pure
quantum state).  In other words, the quantum state would be written into the underlying reality of the world and we could assert that the quantum state is real.  Harrigan and Spekkens
\cite{harrigan2010einstein} introduced the following terminology: $\psi$-epistemic models are those in which there exist pairs of pure quantum states for which the distributions over ontic
states overlap; $\psi$-ontic models are those in which the distributions over ontic states are non-overlapping for any pair of pure quantum states.  In $\psi$-ontic models we can assert that
the wavefunction is a real thing. The question is whether all reasonable models reproducing the predictions of quantum theory are $\psi$-ontic.

There has been a lot of interest in this subject recently.  In 2008 Montina \cite{montina2008exponential, montina2011state} showed that, under the very natural assumption that the evolution
of the $\lambda$ is Markovian, the hidden variables must have at least $2N-2$ real parameters where $N$ is the dimension of the Hilbert space (this being the same number of real parameters
required to describe a pure quantum state).  This provides rather strong evidence for the reality of the quantum state.   Then, in 2011 Pusey, Barrett, and Rudolph (PBR)
\cite{pusey2011quantum} obtained the momentous result that, under a certain separability assumption, distinct pure quantum states have non-overlapping probability distributions thus
providing the first proof that the quantum state is a real thing in the terms outlined above.  The separability assumption employed by PBR is that independently prepared pure quantum states
correspond to a product of probability distributions over ontic states.

In this paper we show that, if three assumptions hold, then pure quantum states have non-overlapping ontic support thus implying the reality of the wavefunction.   The key assumption employed here,
\emph{ontic indifference}, is quite different from the separability assumption used by PBR.  Additionally the proof presented here only requires a single system (whereas PBR required
multiple copies of the system of interest).  The assumption of ontic indifference is that it is possible to implement any quantum transformation which does not affect any given pure quantum
state, $|\varphi\rangle$, in such a way that the ontic states in the support of that state are not affected.

At the cost of very slightly complicating matters we will also show that can run the proof with an even weaker assumption.  This is the assumption of \emph{restricted ontic indifference} in
which it is assumed that there exists at least one pure state, $|0\rangle$, such that quantum transformations leaving this state unchanged can be implemented in a way that does not affect
the ontic states in the ontic support of $|0\rangle$.  An example is the following. Consider a system consisting of a quantum particle that can be in one of $N+1$ boxes labeled $n=0$ to $N$
such that we have an $N+1$ dimensional Hilbert space (we do not consider any internal structure in the boxes).  Let $|0\rangle$ be the state corresponding to the particle being localized in
box 0.  Consider further different possible quantum transformations we might perform that leave $|0\rangle$ invariant.  Then restricted ontic indifference asserts that we can find an
apparatus to implement implement these quantum transformations in such a way that ontic states in the support of $|0\rangle$ are not affected by which quantum transformation we choose to
implement.  Transformations that leave $|0\rangle$ invariant can all be implemented without actually touching box 0.  Indeed, boxes 1 to $N$ could be at a great distance from box 0.  Viewed
in this way restricted ontic indifference is a kind of locality assumption.  Restricted ontic indifference would follow if (i) all the ontic variables associated with a localized particle are \lq\lq situated" in the region where that particle is localized, and (ii)these ontic variables are unaffected by transformations that can be implemented without touching this region.

In fact the assumption of ontic indifference (and of restricted ontic indifference) is violated by the Spekkens toy model mentioned above and, even more interestingly, by a particular application of that model due to Elliott Martin and Robert Spekkens (see \cite{spekkens2008why}) which applies to interferometers of the sort that we will consider later.  In the model of Martin and Spekkens there are ontic states associated with a path of an interferometer even when the particle goes along the other path (i.e. it violates (i) from the preceding paragraph).  This can be viewed as taking a field picture of the ontic variables in which the vacuum has non-trivial structure.  The assumption of restricted ontic indifference is not violated in the de Broglie Bohm model for quantum mechanics in which the ontic state is given by $(\psi, {\bf x})$ where $\bf x$ is the actual position, in configuration space, of the particles.  Thus the result of this paper could be read in the following way.  While it is possible to construct $\psi$-ontic models that satisfy the ontic indifference assumption, it is not possible to build $\psi$-epistemic models of this nature.  This, then, could be regarded as strong hint on how to construct of $\psi$-epistemic models.  In particular, it points to a field (rather than a particle) ontology for $\psi$-epistemic models.  

The main proof we will present requires that the Hilbert space associated with the system in question is infinite dimensional.   If this dimension, $N+1$, is finite, then we prove that any
pair of pure states, $|\varphi\rangle$ and $|\psi\rangle$, for which $\langle \varphi| \psi\rangle|^2 \geq \frac{N-1}{N}$ must have non-overlapping distributions over the ontic states (under the given assumptions).   We
will also discuss the case of a system, $A$, with finite dimensional Hilbert space.  We will show how, by introducing an infinite dimensional ancilla, $B$, in some fixed state, we are able
to deduce that different pure states for system $A$ must correspond to different ontic states of $AB$.

We will introduce the assumptions we use in Sec.\ \ref{assumptions}.   In Sec.\ \ref{interferometer} we will illustrate the proof method using a Mach Zehnder interferometer.  This works for
the $N=2$ case.  Then, in Sec.\ \ref{generalproof}, we will prove the main theorem using the assumption of ontic indifference and assuming that the Hilbert space dimension can be arbitrarily
large.  In Sec.\ \ref{proofwithrestricted} we will show how to replace the ontic indifference assumption with the restricted ontic indifference assumption.  Finally we show how to deal with
systems having a finite dimensional Hilbert space by using an ancilla.

\section{Assumptions} \label{assumptions}

We make the following assumptions \begin{description} \item[Realism.] Each time a system is prepared there exists an underlying state of reality, $\lambda$, which we will call the ontic
state. \item[Possibilistic completeness.]  The ontic state, $\lambda$, is sufficient to determine whether any outcome of any measurement has probability equal to zero of occurring or not.
\item[Ontic indifference.] Any quantum transformation on a system which leaves unchanged any given pure state, $|\psi\rangle$, can be performed in such a way that it does not affect the
    underlying ontic states, $\lambda\in\Lambda_{|\psi\rangle}$, in the ontic support of that pure state.
\end{description} By the \emph{ontic support} of a given state, $|\psi\rangle$, we mean the set, $\Lambda_{|\psi\rangle}$, of ontic states, $\lambda$ which might be prepared when the given
pure state is prepared (i.e.\ those ontic states that have a non-zero probability of being prepared when the given pure state is prepared).

Rather than assuming possibilistic completeness we might have made the stronger assumption that $\lambda$ determines the probability.  However, it is sufficient for our purposes that it only
determines whether the probability is zero or non-zero.  Of course, in any model having hidden variables, $\lambda$, it is reasonable to expect the hidden variables actually determine the probabilities for measurement outcomes.

We assume that the system has arbitrarily large Hilbert space dimension.  We also assume that we can perform arbitrary unitary transformations and arbitrary measurements on the system (or,
at least, we assume that we can perform the particular unitary transformations and measurements that we need to run the proof).  This is reasonable in view of the fact that arbitrary
unitaries can be implemented with beamsplitters and mirrors in the case of interferometry as shown by Reck and Zeilinger \cite{reck1994experimental} and, more generally, we can perform
arbitrary unitaries in quantum circuits given access to a universal gate set \cite{deutsch1985quantum}.  An arbitrary measurement can then be implemented by having the appropriate unitary
followed by a measurement in some standard basis.

Similar assumptions to the first two assumptions appear in the work of Pusey, Barrett, and Rudolph and it is difficult to imagine getting any traction on the problem unless we make these
assumptions (though, of course, it is still interesting to investigate models that do not satisfy these assumptions).  The key assumption here is the third assumption.  It turns out that we can weaken this assumption and still run the proof.  Thus we can, instead, assume
\begin{description}
\item[Restricted ontic indifference] Any quantum transformation on a system which leaves a particular given pure quantum state, $|0\rangle$, unchanged can be implemented in such a way
    that it does not affect the underlying ontic states, $\lambda\in\Lambda_{|0\rangle}$, in the ontic support of $|0\rangle$.
\end{description}
This is a considerably weaker assumption and is very well motivated in the case where $|0\rangle$ corresponds to a spatially localized particle as discussed in Sec.\
\ref{introduction}.

We will first run the proof using the assumption of ontic indifference then show how we can, in fact, use the weaker assumption of restricted ontic indifference.


\section{Interferometric example} \label{interferometer}

\begin{figure}[t]
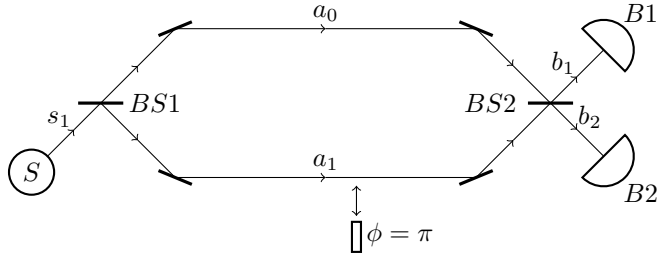
\begin{Diagram}[0.8]{3}{0} \point{S}{45:-5}  \point{BS1}{0,0} \point{MT1}{45:7} \point{MB1}{-45:7}
\point{MT2}{$(45:7)+(20,0)$} \point{MB2}{$(-45:7)+(20,0)$}
                  \point{BS2}{$(45:7)+(20,0)+(-45:7)$} \point{B1}{$(45:7)+(20,0)+(-45:7)+(45:5)$} \point{B2}{$(45:7)+(20,0)+(-45:7)+(-45:5)$} \point{PS}{17,-5.5}
\source[45+180]{S} \symb{S} \awire{S}{BS1} \symb[-1,0.7]{s_1} \awire{BS1}{MT1} \awire{BS1}{MB1} \awire{MT1}{MT2} \symb[0,1]{a_0} \awire{MB1}{MB2} \symb[0,1]{a_1} \awire{MT2}{BS2}
\awire{MB2}{BS2} \awire{BS2}{B1} \symb[135:1.3]{b_1}    \awire{BS2}{B2}  \symb[45:1.2]{b_2} \detector[-45]{B1}  \symb[45:3.5]{B1} \detector[+45+180]{B2} \symb[-45:3.5]{B2}
\beamsplitter{BS1}\symb[3.6,0]{BS1}  \beamsplitter{BS2} \symb[-4,0]{BS2} \mirror[22.5]{MT1} \mirror[-22.5]{MT2} \mirror[-22.5]{MB1} \mirror[22.5]{MB2} \phaseshifter[-90]{PS}
\symb[2.9,-0.35]{\phi=\pi}
\end{Diagram}
\caption{\label{interferometerone} This apparatus prepares $|\psi\rangle=\frac{1}{\sqrt{2}}(|a_0\rangle+|a_1\rangle)$ which then impinges on \emph{BS2} followed by two
detectors \emph{B1} and \emph{B2}.  A $\phi=\pi$ phase shifter can be placed in path $a_1$.} \end{figure}

In popular accounts of quantum theory an interferometer is often used to argue that something can be in two places at once.  For example, consider the Mach Zehnder interferometer in Fig.\
\ref{interferometerone}.
 The two beamsplitters, \emph{BS1} and \emph{BS2} are 50:50.  We assume that the
transformation effected at each beamsplitter is given by the unitary matrix \begin{equation} \frac{1}{\sqrt{2}} \left( \begin{array}{cc} 1 & 1 \\
                        -1 & 1  \end{array} \right)
\end{equation}
Such that
\begin{equation} |s_1\rangle \longrightarrow \frac{1}{\sqrt{2}} ( |a_0\rangle + |a_1\rangle ) \end{equation}
at \emph{BS1} and
\begin{equation} |a_0\rangle
\longrightarrow \frac{1}{\sqrt{2}} (|b_1\rangle + |b_2\rangle )  ~~~~ |a_1\rangle \longrightarrow \frac{1}{\sqrt{2}} (|b_1\rangle - |b_2\rangle ) \end{equation}
at \emph{BS2}.  We have
chosen the integer subscript labeling here for later convenience.  We require, of course, that \begin{equation} \langle s_m|s_n\rangle = \delta_{mn}  ~~~~~ \langle a_m|a_n\rangle =
\delta_{mn} ~~~~~ \langle b_m|b_n\rangle = \delta_{mn} \end{equation} We will assume that the path lengths are such that zero net phase is accumulated as the particle goes along the internal
paths of the interferometer.  On passing through the first, then the second beamsplitter it follows from the above transformations that the evolution is \begin{equation} |s_1\rangle
\longrightarrow \frac{1}{\sqrt{2}} ( |a_0\rangle + |a_1\rangle ) \longrightarrow |b_1\rangle \end{equation} Thus the detector, $B_1$, in path $b_1$ will always fire and there is zero
probability that the $B_2$ detector will fire. However, if we insert a phase shifter $\phi=\pi$ in the $a_2$ path as indicated then the evolution is \begin{equation} |s_1\rangle
\longrightarrow \frac{1}{\sqrt{2}} ( |a_0\rangle + |a_1\rangle ) \longrightarrow \frac{1}{\sqrt{2}} ( |a_0\rangle - |a_1\rangle )\longrightarrow |b_2\rangle \end{equation} With the phase
shifter in place the detector $B_2$ will always fire and there is zero probability that the detector $B_1$ will fire.

Given this set up, one version of the popular argument for something going both ways is the following.  Let us try to account for this experiment in such a way that the particle either goes
along path $a_0$ or $a_1$.  If this is the case then, sometimes it must go along path $a_0$.  Consider a case in which it does go along path $a_0$.  When the particle reaches \emph{BS2} it
must go along path $b_1$ if there is no phase shifter and it must go along path $b_2$ if the $\phi=\pi$ phase shifter is in place. Since it did not go along path $a_1$ it does not \lq\lq
know" whether there is a phase shifter in place or not.  Hence, it cannot make the correct \lq\lq decision".  This shows that something must be going along both paths.  We will now see how
to turn this argument into a formal argument for the reality of the quantum state.

\begin{figure}
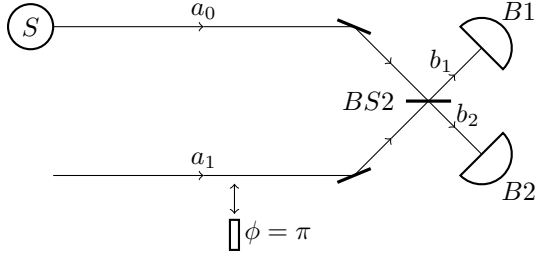
 \begin{Diagram}[0.8]{3}{0} \point{S}{45:-5}  \point{BS1}{0,0} \point{MT1}{45:7} \point{MB1}{-45:7}
\point{MT2}{$(45:7)+(20,0)$} \point{MB2}{$(-45:7)+(20,0)$}
                  \point{BS2}{$(45:7)+(20,0)+(-45:7)$} \point{B1}{$(45:7)+(20,0)+(-45:7)+(45:5)$} \point{B2}{$(45:7)+(20,0)+(-45:7)+(-45:5)$}
                  \point{PS}{17,-5.5}
\source[180]{MT1} \symb{S} \awire{MT1}{MT2} \symb[0,1]{a_0} \awire{MB1}{MB2} \symb[0,1]{a_1} \awire{MT2}{BS2} \awire{MB2}{BS2} \awire{BS2}{B1} \symb[135:1.3]{b_1}    \awire{BS2}{B2}
\symb[45:1.2]{b_2} \detector[-45]{B1}  \symb[45:3.5]{B1} \detector[+45+180]{B2} \symb[-45:3.5]{B2}
  \beamsplitter{BS2} \symb[-4,0]{BS2}  \mirror[-22.5]{MT2} \mirror[22.5]{MB2}
\phaseshifter[-90]{PS} \symb[2.9,-0.35]{\phi=\pi} \end{Diagram} \caption{\label{interferometertwo} This apparatus prepares the state $|\varphi\rangle=|a_0\rangle$ which then impinges on \emph{BS2} followed by two
detectors \emph{B1} and \emph{B2}. A  $\phi=\pi$ phase shifter can be inserted in path $a_1$.} \end{figure}

This example can be used to show that the pair of states
\begin{equation} |\varphi\rangle = |a_1\rangle ~~~~\text{and} ~~~~~ |\psi\rangle=\frac{1}{\sqrt{2}}(|a_0\rangle+ |a_1\rangle)
\end{equation}
must have non-overlapping ontic support given three assumptions in Sec.\ \ref{assumptions} above.  We can prepare the state $|\varphi\rangle=|a_0\rangle$ by the apparatus
shown in Fig.\ \ref{interferometertwo} and we can prepare the state $|\psi\rangle=\frac{1}{\sqrt{2}}(|a_0\rangle+ |a_1\rangle)$ by the apparatus shown in Fig.\ \ref{interferometerone} (in each case we consider the state at a time just before the system passes through the location where the phase shifter may be inserted).
 Let $\Lambda_{|\varphi\rangle}$ be the ontic support of the state
$|\varphi\rangle=|a_0\rangle$.   We now allow it to impinge on the measurement apparatus constituted by \emph{BS2} and the two detectors $B_1$ and $B_2$.  Further, we have the phase shifter,
$\phi$, which can be inserted in path $a_1$. We have a choice of two setting for the phase shifter.  Either $\phi=0$ (not inserted) or $\phi=\pi$ (inserted). Let
$\Lambda_{|\varphi\rangle}^{B_1}[\phi] \subseteq \Lambda_{|\varphi\rangle}$ be the subset of ontic states in the ontic support of $|\varphi\rangle$ that have non-zero probability of giving
rise to a click at detector $B_1$ given that the phase shifter in path $a_1$ has setting $\phi$.  We are invoking possibilistic completeness in asserting that this set is well defined.
Likewise we define $\Lambda_{|\varphi\rangle}^{B_2}[\phi] \subseteq \Lambda_{|\varphi\rangle}$ to be the subset of ontic states in the ontic support of $|\varphi\rangle$ that have non-zero
probability of giving rise to a click at detector $B_2$ with phase shifter setting $\phi$.  Inserting, or not inserting, the phase shifter has no affect on the quantum state
$|\varphi\rangle=|a_0\rangle$.  Hence, it follows from the assumption of ontic indifference that inserting, or not inserting, the phase shifter has no affect on the ontic states, $\lambda\in
\Lambda_{|\varphi\rangle}$ in the support of $|\varphi\rangle$.
It follows that
\begin{eqnarray} \Lambda_{|\varphi\rangle}^{B_1}[\phi=0] =\Lambda_{|\varphi\rangle}^{B_1}[\phi=\pi] =
\Lambda_{|\varphi\rangle}^{B_1}\\ \Lambda_{|\varphi\rangle}^{B_2}[\phi=0] =\Lambda_{|\varphi\rangle}^{B_2}[\phi=\pi] = \Lambda_{|\varphi\rangle}^{B_2}            \label{onticindifB2}
\end{eqnarray} We have \begin{equation} \label{B1cupB2} \Lambda_{|\varphi\rangle}^{B_1} \cup \Lambda_{|\varphi\rangle}^{B_2} = \Lambda_{|\varphi\rangle} \end{equation}
as either detector $B_1$ or $B_2$ must fire (we are assuming 100\% efficient detectors).

Now consider an ontic state, $\lambda$, in the intersection \begin{equation} \Lambda_{|\varphi\rangle}^{B_2}[\phi=0]\cap \Lambda_{|\psi\rangle} \end{equation} where $\Lambda_{|\psi\rangle}$
is the ontic support of $|\psi\rangle=\frac{1}{\sqrt{2}}(|a_0\rangle+ |a_1\rangle)$.    If $\phi=0$ and $|\psi\rangle$ has been prepared then we know that detector $B_2$ cannot fire.  Hence,
using (\ref{onticindifB2}), \begin{equation} \Lambda_{|\varphi\rangle}^{B_2}[\phi=0]\cap \Lambda_{|\psi\rangle} = \Lambda_{|\varphi\rangle}^{B_2}\cap \Lambda_{|\psi\rangle} = \emptyset
\end{equation} (where $\emptyset$ is the empty set) since $\Lambda_{|\varphi\rangle}^{B_2}[\phi=0]$ are the ontic states that have non-zero probability of causing $B_2$ to fire.  Likewise,
if we consider the setting $\phi=\pi$ then we know that detector $B_1$ cannot fire and hence \begin{equation} \Lambda_{|\varphi\rangle}^{B_1}\cap \Lambda_{|\psi\rangle} = \emptyset
\end{equation} Using (\ref{B1cupB2}) this implies that \begin{equation}\label{motivatingonticdistinct} \Lambda_{|\varphi\rangle}\cap \Lambda_{|\psi\rangle} = \emptyset \end{equation} since
\begin{equation} \Lambda_{|\varphi\rangle}\cap \Lambda_{|\psi\rangle} = (\Lambda_{|\varphi\rangle}^{B_1} \cup \Lambda_{|\varphi\rangle}^{B_2}) \cap \Lambda_{|\psi\rangle} =
(\Lambda_{|\varphi\rangle}^{B_1} \cap \Lambda_{|\psi\rangle}) \cup (\Lambda_{|\varphi\rangle}^{B_2} \cap \Lambda_{|\psi\rangle}) = \emptyset \end{equation} We see here the crucial role the
assumption of ontic indifference is playing since if we had to keep the $\phi$ argument on $\Lambda_{|\varphi\rangle}^{B_1}$ and $\Lambda_{|\varphi\rangle}^{B_2}$ we could not use the two
different contexts in the same equation.   It follows from (\ref{motivatingonticdistinct}) that the states $|\varphi\rangle$ and $|\psi\rangle$ have non-overlapping ontic support.   This
particular proof applies only to a pair of states having $|\langle \varphi|\psi\rangle|^2=\frac{1}{2}$.  We would like to prove that any pair of distinct pure states has non-overlapping
ontic support.

\begin{figure}[t]
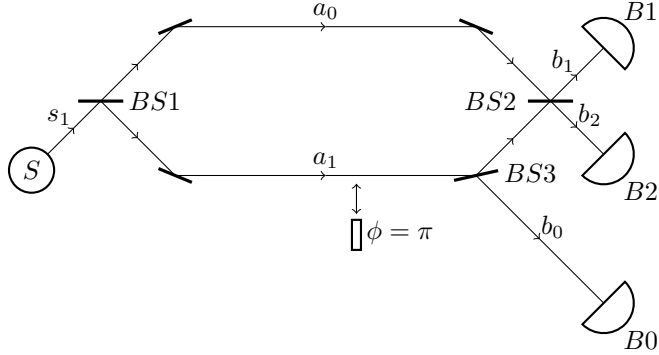
 \begin{Diagram}[0.8]{3}{0} \point{S}{45:-5}  \point{BS1}{0,0} \point{MT1}{45:7} \point{MB1}{-45:7}
\point{MT2}{$(45:7)+(20,0)$} \point{MB2}{$(-45:7)+(20,0)$}
                  \point{BS2}{$(45:7)+(20,0)+(-45:7)$} \point{B1}{$(45:7)+(20,0)+(-45:7)+(45:5)$} \point{B2}{$(45:7)+(20,0)+(-45:7)+(-45:5)$}
                  \point{B0}{$(-45:7)+(20,0)+(-45:12)$}
                  \point{PS}{17,-5.5}
\source[45+180]{S} \symb{S} \awire{S}{BS1} \symb[-1,0.7]{s_1} \awire{BS1}{MT1} \awire{BS1}{MB1} \awire{MT1}{MT2} \symb[0,1]{a_0} \awire{MB1}{MB2} \symb[0,1]{a_1} \awire{MT2}{BS2}
\awire{MB2}{BS2} \awire{BS2}{B1} \symb[135:1.3]{b_1}    \awire{BS2}{B2}  \symb[45:1.2]{b_2} \detector[-45]{B1}  \symb[45:3.5]{B1} \detector[+45+180]{B2} \symb[-45:3.5]{B2}
\beamsplitter{BS1}\symb[3.6,0]{BS1}  \beamsplitter{BS2} \symb[-4,0]{BS2} \mirror[22.5]{MT1} \mirror[-22.5]{MT2} \mirror[-22.5]{MB1} \beamsplitter[12.5]{MB2} \symb[3.6,0.2]{BS3}
\awire{MB2}{B0}  \symb[45:1.2]{b_0} \detector[45+180]{B0} \symb[-45:3.5]{B0} \phaseshifter[-90]{PS} \symb[2.9,-0.35]{\phi=\pi} \end{Diagram} \caption{\label{interferometerthree} This apparatus prepares the state
$|\psi\rangle= \alpha |a_0\rangle + \beta |a_1\rangle$ which then impinges onto \emph{BS2} and \emph{BS3} and the three detectors \emph{B0}, \emph{B1}, \emph{B2}.  A $\phi=\pi$ phase shifter
can be placed in path $a_1$. } \end{figure}

\begin{figure}[t]
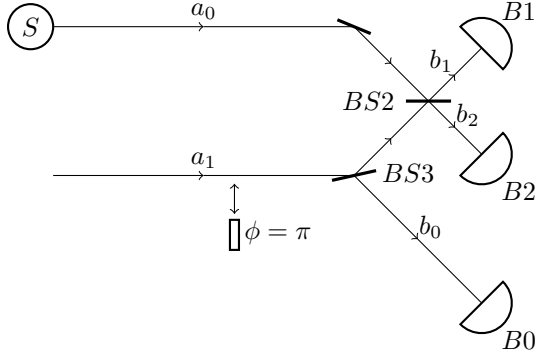
 \begin{Diagram}[0.8]{3}{0} \point{S}{45:-5}  \point{BS1}{0,0} \point{MT1}{45:7} \point{MB1}{-45:7}  \point{MT2}{$(45:7)+(20,0)$}
\point{MB2}{$(-45:7)+(20,0)$}
                  \point{BS2}{$(45:7)+(20,0)+(-45:7)$} \point{B1}{$(45:7)+(20,0)+(-45:7)+(45:5)$} \point{B2}{$(45:7)+(20,0)+(-45:7)+(-45:5)$}
                  \point{B0}{$(-45:7)+(20,0)+(-45:12)$}
                  \point{PS}{17,-5.5}
\source[180]{MT1} \symb{S}    \awire{MT1}{MT2} \symb[0,1]{a_0} \awire{MB1}{MB2} \symb[0,1]{a_1} \awire{MT2}{BS2} \awire{MB2}{BS2} \awire{BS2}{B1} \symb[135:1.3]{b_1}    \awire{BS2}{B2}
\symb[45:1.2]{b_2} \detector[-45]{B1}  \symb[45:3.5]{B1} \detector[+45+180]{B2} \symb[-45:3.5]{B2} \beamsplitter{BS2} \symb[-4,0]{BS2}  \mirror[-22.5]{MT2}\beamsplitter[12.5]{MB2}
\symb[3.6,0.2]{BS3}   \awire{MB2}{B0} \symb[45:1.2]{b_0}
 \detector[45+180]{B0} \symb[-45:3.5]{B0}
\phaseshifter[-90]{PS} \symb[2.9,-0.35]{\phi=\pi} \end{Diagram} \caption{\label{interferometerfour} This apparatus prepares the state $|\varphi\rangle= |a_0\rangle$ which then impinges onto \emph{BS2} and \emph{BS3}
and the three detectors \emph{B0}, \emph{B1}, \emph{B2}.  A $\phi=\pi$ phase shifter can be placed in path $a_1$. } \end{figure}

Before we prove this we notice that the above proof can be modified to apply to a pair of states having $|\langle \varphi|\psi\rangle|^2 \leq\frac{1}{2}$.   Consider the experiment in Fig.\
\ref{interferometerthree}.  Now assume that the first beamsplitter, \emph{BS1}, is uneven such that it prepares a state $|\psi\rangle=\alpha|a_0\rangle + \beta |a_1\rangle $
where $\alpha$ and $\beta$ are real and positive and  $\alpha\leq\beta$. After the phase shifter the state will impinge on the beamsplitter \emph{BS3} in path $a_1$. We set the
transmissivity, $T$, of this beamsplitter such that \begin{equation} |\psi\rangle = \alpha|a_0\rangle + \beta|a_1\rangle \longrightarrow \alpha|a_0\rangle +\alpha|a'_1\rangle +
\gamma|b_0\rangle \end{equation} This is achieved if $\sqrt{T}\beta = \alpha$.  Note that if the phase shifter had been inserted there would be a minus sign in front of the second term on
the right.  Rather than sending in the state $|\psi\rangle$, we can send in the state $|\varphi\rangle=|a_0\rangle$ as shown in Fig.\ \ref{interferometerfour}.
 Since the amplitudes in front of the $|a_0\rangle$ and
$|a'_1\rangle$ terms are now equal we have probability zero of getting a detection in the $B_2$ detector if $\phi=0$ (phase shifter not inserted) and probability zero of getting a detection
in the $B_1$ detector if $\phi=\pi$ (phase shifter inserted). We define $\Lambda_{|\varphi\rangle}^{B_n}[\phi]$ as being the subsets of $\Lambda_{|\varphi\rangle}$ which have non-zero
probability of giving rise to a detector click in $B_n$ (where $n=0, 1, 2$) when the phase shifter is set to $\phi$.  By ontic indifference, we can  put \begin{equation}
\Lambda_{|\varphi\rangle}^{B_n}[\phi]=\Lambda_{|\psi\rangle}^{B_n} \end{equation} We have that \begin{equation} \Lambda_{|\varphi\rangle}^{B_0} \cup \Lambda_{|\varphi\rangle}^{B_1} \cup
\Lambda_{|\varphi\rangle}^{B_2} = \Lambda_{|\varphi\rangle} \end{equation} Now, we know that the probability of a click at detector $B_0$ given that the state was $|\varphi\rangle$ is zero.
Hence, \begin{equation} \Lambda_{|\varphi\rangle}^{B_0} = \emptyset \end{equation} Also, by the same reasoning as before, we have that \begin{equation} \Lambda_{|\varphi\rangle}^{B_1}\cap
\Lambda_{|\psi\rangle} = \emptyset ~~~~~~ \Lambda_{|\varphi\rangle}^{B_2}\cap \Lambda_{|\psi\rangle} = \emptyset \end{equation} It follows from these equations that \begin{flalign}
\Lambda_{|\varphi\rangle}\cap \Lambda_{|\psi\rangle} &= (\Lambda_{|\varphi\rangle}^{B_0} \cup \Lambda_{|\varphi\rangle}^{B_1} \cup \Lambda_{|\varphi\rangle}^{B_2}) \cap
\Lambda_{|\psi\rangle} \nonumber \\ &= (\Lambda_{|\varphi\rangle}^{B_0}\cap \Lambda_{|\psi\rangle})\cup (\Lambda_{|\varphi\rangle}^{B_1}\cap \Lambda_{|\psi\rangle}) \cup
(\Lambda_{|\varphi\rangle}^{B_2}\cap \Lambda_{|\psi\rangle}) = \emptyset \end{flalign} Since we can choose transmissivity, $T$, for \emph{BS3} such that $\sqrt{T}\beta =\alpha$ whenever
$\beta \geq \frac{1}{\sqrt{2}}$ it follows that we can argue that the ontic support of the states $|\varphi\rangle$ and $|\psi\rangle$ are non-overlapping whenever $|\langle
\varphi|\psi\rangle|^2 \leq\frac{1}{2}$.

\section{General proof}\label{generalproof}

The above examples were illustrated with interferometers.  Now, for the sake of generality, we will simply consider abstract Hilbert spaces without specifying a particular physical
instantiation so that the argument is fully general. Consider a $N+1$ dimensional Hilbert space.  We will need to allow $N$ to be arbitrarily large to run the proof (we will discuss ways
around this later).  We wish to show that any pair of distinct pure states, $|\varphi\rangle$ and $|\psi\rangle$, have non-overlapping ontic support.  We can write \begin{equation}
|\varphi\rangle = |a_0\rangle   ~~~~~~ |\psi\rangle = \alpha |a_0\rangle + \beta |a_1\rangle \end{equation} for some orthonormal basis $\{|a_n\rangle: n=0 ~\text{to}~ N\}$ (having $\langle
a_m|a_n\rangle = \delta_{mn}$).  We take $\alpha$ and $\beta$ to be real and positive (we can always do this since we can absorb any phase into the definition of $|a_0\rangle$ and
$|a_1\rangle$).

We consider unitary transformations which leave $|\varphi\rangle$ unchanged (so that we we can use the ontic indifference assumption).  Such transformations take the form \begin{equation} U=
\left( \begin{array}{c|ccc} 1 & 0  & 0 & \hdots  \\ \hline 0 & {} & {}& {}   \\ 0 & {} & V & {}  \\ \vdots &{}&{}&{} \end{array} \right) \end{equation} in the $|a_n\rangle$ basis where $V$
is any unitary matrix acting in the subspace spanned by $|a_1\rangle$ to $|a_N\rangle$.  In particular, we will consider the set of such transformations \begin{equation}\label{Umform} U[m]=
\left( \begin{array}{c|ccc} 1 & 0  & 0 & \hdots  \\ \hline 0 & {} & {}& {}   \\ 0 & {} & V[m] & {}  \\ \vdots &{}&{}&{} \end{array} \right) \end{equation} for some choice of $V[m]$. All of
these leave the state $|\varphi\rangle$ unchanged and hence can be implemented in such a way that they leave ontic states $\lambda\in\Lambda_{|\varphi\rangle}$ (in the ontic support of
$|\varphi\rangle$) unchanged by the ontic indifference assumption (we will always assume that such transformations are implemented in such a way).

After the transformation $U[m]$ we have \begin{equation} |\psi\rangle = \alpha |a_0\rangle + \beta |a_1\rangle \longrightarrow \alpha |a_0\rangle + \beta |b[m]\rangle \end{equation} where
$|b[m]\rangle = V[m] |a_1\rangle$ is normalized and orthogonal to $|a_0\rangle$.  By appropriate choice of $V[m]$, the $|b[m]\rangle$ can be equal to any normalized vectors in the subspace
orthogonal to $|a_0\rangle$. The $|b[m]\rangle$ vectors do not have to be orthogonal to one another.  We put \begin{equation} |c[m]\rangle =  \alpha |a_0\rangle + \beta |b[m]\rangle
\end{equation} The states $|c[m]\rangle$ are normalized but not necessarily orthogonal to each other.

Following the transformation $U[m]$ we subject the system to a maximal measurement (i.e.\ a measurement corresponding to a non-degenerate observable) onto a basis $\{|d_n\rangle: n=0~
\text{to}~N\}$ (we denote by $D_n$ the measurement outcome associated with the basis vector $|d_n\rangle$).  We wish to choose this measurement and the $V[n]$ such that:
\begin{equation}\label{maincondition} \text{For all}~ n ~\text{EITHER} ~~~~ \langle d_n|a_0\rangle=0 ~~~~\text{OR}~~~~ \langle d_n | c[n] \rangle = 0 \end{equation} (the OR is not exclusive
- it is ok if both conditions are satisfied). First we will show that this condition can be satisfied for any situation where $\beta > 0$ (so $|\varphi\rangle$ and $|\psi\rangle$ are
distinct) and then we will show that it implies that disjoint pure states have non-overlapping ontic support.

We will prove that we can actually satisfy the condition (\ref{maincondition}).  We will work in the $\{|d_n\rangle:n=1~\text{to} N\}$ basis.  In this basis we put
\begin{equation}\label{a0choice} |a_0\rangle = \frac{1}{\sqrt{M}}(0, \underbrace{1, 1, \dots, 1}_M, 0, 0,\dots, 0) \end{equation} where $M$ is the smallest integer such that
\begin{equation}\label{conditiononM} M \geq \frac{1}{\beta^2} \end{equation} (we require $M\leq N$).  Since $\alpha^2+\beta^2=1$ this is equivalent to \begin{equation}\label{conditiononM2}
\frac{\alpha^2}{\beta^2} \leq M-1 \end{equation} We are free to choose the $|b[m]\rangle$ states to be any normalized states orthogonal to $|a_0\rangle$. We choose
\begin{equation}\label{bmprechoice} |b[m]\rangle = \frac{1}{\sqrt{M-1}} \frac{\alpha}{\beta} |\overline{b}[m]\rangle + \delta |d_0\rangle \end{equation} where $|\overline{b}[m]\rangle$ is
orthogonal to $|d_0\rangle$.  It follows from (\ref{conditiononM2}) that the first coefficient on the right hand side is less than one.  We choose the coefficient $\delta$ to guarantee the
state is normalized.  It follows that \begin{equation}\label{cmprechoice} |c[m]\rangle = \alpha  (|a_0\rangle + \frac{1}{\sqrt{M-1}}|\overline{b}[m]\rangle) + \gamma |d_0\rangle
\end{equation} where $\gamma=\beta\delta$.  We wish to choose these states so that (\ref{maincondition}) is satisfied.  One way of doing this is if we can arrange things so that, in the
$\{|d_n\rangle:n=1~\text{to} N\}$ basis, \begin{equation}\label{cmchoice} |c[m]\rangle = \frac{1}{\sqrt{M-1}}(0,\underbrace{1, 1, \dots, 1, \overbrace{0}^m, 1, \dots ,1}_M, 0, 0, \dots, 0)
\end{equation} for $m=1$ to $M$.  Then $\langle d_n|c[n]\rangle=0$ for $n=1$ to $M$.  For $m=0$ and $m=M+1$ to $N$ we put $U[m]=U[1]$.  It follows that $\langle d_n|a_0\rangle = 0$ for $n=0$
and $n=M+1$ to $N$.   Hence, if we can choose $|c[m]\rangle$ as in (\ref{cmchoice}) then condition (\ref{maincondition}) is satisfied.  To see that we can cho0se $|c[m]\rangle$  in this way
we note that it follows from (\ref{a0choice}), (\ref{cmprechoice}), and (\ref{cmchoice}) after a little algebra that \begin{equation} \langle a_0| b[m] \rangle = 0 ~~~ \text{for} ~m=0
~\text{to}~N \end{equation} (indeed, the choice in (\ref{bmprechoice}) was made to make this work).  This is the only condition that has to be satisfied - it ensures that we can find $U[m]$
of the form given in (\ref{Umform}) such that $|c[m]\rangle=U[m]|\psi\rangle$.

Now we will show that it follows from (\ref{maincondition}) that $|\varphi\rangle$ and $|\psi\rangle$ have non-overlapping ontic support (as long as they are distinct).  Let
$\Lambda_{|\varphi\rangle}^{D_n}[m]$ be the subset of the ontic support of $|\varphi\rangle$ whose members have a non-zero probability of giving rise to measurement outcome $D_n$ when we
precede this measurement by the unitary $U[m]$.  By ontic indifference, \begin{equation} \Lambda_{|\varphi\rangle}^{D_n}[m] = \Lambda_{|\varphi\rangle}^{D_n}[m'] =
\Lambda_{|\varphi\rangle}^{D_n}  ~~~~ \text{for all}~m, m' \end{equation} since the state $|\varphi\rangle$ is unaffected by $U[m]$. We also have \begin{equation} \Lambda_{|\varphi\rangle} =
\bigcup_{n=0}^N \Lambda_{|\varphi\rangle}^{D_n} \end{equation} since we must see some outcome for the measurement. Assume that we perform the unitary $U[n]$ and then measure in the
$\{|d_n\rangle\}$ basis. Then \begin{equation}\label{LambdaDnpsiempty}
 \Lambda_{|\varphi\rangle}^{D_n} \cup \Lambda_{|\psi\rangle} = \emptyset
\end{equation} This is because (using (\ref{maincondition})) EITHER $\langle d_n|\psi\rangle=\langle d_n|a_0\rangle=0$ in which we must have case $\Lambda_{|\varphi\rangle}^{D_n}=\emptyset$
from which (\ref{LambdaDnpsiempty}) follows, OR  $\langle d_n| U[n]| \psi\rangle = \langle d_n |c[n] \rangle = 0$ in which case (\ref{LambdaDnpsiempty}) follows since there can be no ontic
states which both have non-zero probability of giving rise to outcome $D_n$ (since they belong to $\Lambda_{|\varphi\rangle}^{D_n}$) and have zero probability of giving rise to this same
outcome (as they belong to $\Lambda_{|\psi\rangle}$).  It then follows that \begin{equation}
 \Lambda_{|\varphi\rangle} \cap \Lambda_{|\psi\rangle} = \left( \bigcup_{n=0}^N \Lambda_{|\varphi\rangle}^{D_n} \right) \cap  \Lambda_{|\psi\rangle}
 =  \bigcup_{n=0}^N \left( \Lambda_{|\varphi\rangle}^{D_n} \cap  \Lambda_{|\psi\rangle} \right) = \emptyset
\end{equation} Hence the ontic supports of distinct states are non-overlapping.  As long as the Hilbert space dimension, $N+1$, of the system in question is large enough then however small
$\beta$ is, we can always find a $M\leq N$ such that the condition in (\ref{conditiononM}) is satisfied.  Hence, in the limit of $N\rightarrow \infty$, any pair of pure states must have
non-overlapping ontic support and therefore the quantum state is a real thing.  In the case that we are restricted to finite dimensional Hilbert spaces it follows from the present argument
using (\ref{conditiononM}) that any pair of states for which \begin{equation} \langle \varphi| \psi\rangle|^2 \geq \frac{N-1}{N} \end{equation} have non-overlapping ontic support.

\section{Proof with restricted ontic indifference} \label{proofwithrestricted}

As before we take the distinct pure states $|\varphi\rangle$ and $|\psi\rangle$.  Previously we subjected these states to a transformation $U[m]$ then performed a measurement in the
$\{|d_n\rangle\}$ basis.  Now we will complicate the experiment a little.  First we subject these states to a unitary transformation $W$ chosen such that $W|\varphi\rangle = |0\rangle$.
Next, at this intermediate time, we perform the transformation $\tilde{U}[m]= WU[m]W^\dagger$.  Then we perform the transformation $W^\dagger$.   Finally we perform a measurement in the
$\{|d_n\rangle\}$ basis.   In this new configuration we can consider the set $\Lambda_{|\varphi\rangle}^{D_n}[m] \subseteq \Lambda_{|\varphi\rangle}$ of ontic states in the support of
$|\varphi\rangle$ at the initial time that have non-zero probability of giving rise to measurement output $D_n$ later given that we are to implement the transformation $\tilde{U}[m]$ at the
intermediate time.    We can also consider the set of ontic states, $\Lambda_{|0\rangle}^{D_n}[m] \subseteq \Lambda_{|0\rangle}$ at the intermediate time which are in the ontic support of
$|0\rangle$ and have a non-zero probability of giving rise to measurement outcome $D_n$  given that we implement transformation $\tilde{U}[m]$ at the intermediate time.  Now it must be the
case that \begin{equation} \Lambda_{|\varphi\rangle} \longrightarrow \Lambda_{|0\rangle} \end{equation} under the evolution $W$.   Hence, it must be true that \begin{equation}
\Lambda_{|\varphi\rangle}^{D_n}[m] \longrightarrow \Lambda_{|0\rangle}^{D_n}[m] \end{equation} as the set on the right is just the time evolved version of the set on the left.  By restricted
ontic indifference we have \begin{equation} \Lambda_{|0\rangle}^{D_n}[m] = \Lambda_{|0\rangle}^{D_n}[m'] = \Lambda_{|0\rangle}^{D_n} \end{equation} Hence \begin{equation}
\Lambda_{|\varphi\rangle}^{D_n}[m] = \Lambda_{|\varphi\rangle}^{D_n}[m'] = \Lambda_{|\varphi\rangle}^{D_n} \end{equation} Thus we have the necessary property to run the proof as in the
previous section.

\section{Systems with finite Hilbert space dimension}\label{finitedimcase}

Imagine that the system, $A$, under consideration necessarily has a finite Hilbert space dimension, $L$.  Can we still argue that its pure states are real?  One way to do this is to allow
the use of an ancillary system, $B$, having arbitrarily large $N$ prepared in some initial state $|1\rangle$.  We can ask whether $|\varphi\rangle_A$ and $|\psi\rangle_A$ for system $A$ must
have non-overlapping ontic support.  Then the states of the joint system under consideration are $|\varphi\rangle_A |1\rangle_B$ and $|\psi\rangle_A|1\rangle_B$.  We can now argue as before
that these states must have non-overlapping ontic support.  That is 
\begin{equation}\label{lambdaAB} \Lambda_{|\varphi\rangle_A |1\rangle_B} \cap \Lambda_{|\psi\rangle_A|1\rangle_B} =
\emptyset \end{equation} 
for any pair of states $|\varphi\rangle_A$ and $|\psi\rangle_A$.  Assume that, in the case for independently prepared pure quantum states, we have
\begin{equation}\label{PBRish} \Lambda_{|\psi\rangle_A|\vartheta\rangle_B} = \Lambda_{|\psi\rangle_A}\times\Lambda_{|\vartheta\rangle_B} \end{equation} 
for any states $|\psi\rangle_A$ and
$|\vartheta\rangle_B$ for systems $A$ and $B$ respectively. Here the $\times$ represents the cartesian product of the sets. We will call this assumption \emph{ontic product separability}.
It follows from (\ref{lambdaAB}) and (\ref{PBRish}) that 
\begin{equation} \Lambda_{|\varphi\rangle_A} \cap \Lambda_{|\psi\rangle_A} = \emptyset \end{equation} 
This implies that any pair of
pure states for the finite dimensional Hilbert space must have non-overlapping ontic support.  The ontic product separability assumption in (\ref{PBRish}) is very natural if we admit the
possibility of considering subsystems of the universe (such as $A$) in isolation from the rest of the universe because then we would want to give the system its own ontic states.   The assumption in (\ref{PBRish}) is part of the PBR separability assumption
\cite{pusey2011quantum}.  If we do not make the ontic product separability assumption then we can still make a strong assertion based on (\ref{lambdaAB}).  Assuming that system $B$ is simply
the rest of the universe it  follows from this equation that the universal ontic situation in which we prepare $|\varphi\rangle_A$ for subsystem $A$ is distinct from the universal ontic
situation in which we prepare $B$.  In the case were we do not assume ontic product separability this is as strong a result as we can expect.

\section{Violating ontic indifference}

In Spekkens's toy model a toy bit is modeled by a system having four underlying ontic states.  States in the theory correspond to having equal probability of two of these four ontic states.  We can imagine a state in the theory which is a equal distribution over the first two ontic states (i.e.\ with equal probabilities).  This can be regarded as being analogous to the $|0\rangle$ state in quantum theory.  Another state would be an equal distribution over the second and third ontic state which would be analogous to a superposition $\frac{1}{\sqrt{2}}(|0\rangle + |1\rangle)$.  An equal distribution over the first and fourth ontic state would be analogous to $\frac{1}{\sqrt{2}}(|0\rangle - |1\rangle)$ in quantum theory.  Measurements course grain over pairs of ontic states (and remix the underlying states so it is impossible to prepare a given ontic state with probability one).  Transformations correspond to permutations of the underlying ontic states.  Thus, we can perform a transformation that interchanges the first and second, and the third and fourth ontic states.  This leaves the state analogous to $|0\rangle$ unchanged but transforms the state analogous to $\frac{1}{\sqrt{2}}(|0\rangle + |1\rangle)$ to the state analogous to $\frac{1}{\sqrt{2}}(|0\rangle - |1\rangle)$.  It does so by interchanging the ontic states in the support of the state analogous to $|0\rangle$.  Thus, ontic indifference is violated in the Spekkens toy model.

An application of this model, due to Martin and Spekkens \cite{spekkens2012why}, pertains to interferometers of the sort ilustrated in Fig.\ \ref{interferometerone} and Fig.\ \ref{interferometertwo}.  In this case there are ontic variables associated with the occupation number of the path (take this to be 0 or 1) and a phase associated with the path (take this to be 0 or $\pi$).   Even if the occupation number is 0 there is still the phase variable which will be affected by a phase shifter.  Thus a path with no particle in it still has non-trivial degrees of freedom associated with it.  This allows the model to violate ontic indifference in a local way.

\section{Previous work and discussion}

In this paper we have been interested in theories that attempt to provide a description of the underlying reality of the world.   This is sometimes referred to as providing an ontology and
sometimes referred to as providing a hidden variable interpretation.  It is only in the context of such theories it is natural to ask whether wave functions are real things.  If we deny that
there is an underlying reality (assuming this is a consistent thing to do) then it does not make sense to ask whether the wave function is a real thing.

The pilot wave model of de Broglie (see \cite{bacciagaluppi2008quantum}) and Bohm \cite{bohm1952suggested} is an interesting example of a theory that attempts to provide a description of the
underlying reality.  In the usual interpretation of this model the ontic state is normally taken to be given by $\lambda = (\psi, \bf{X})$ where $\psi$ is the wavefunction and $\bf X$ is the
position, in configuration space, of the particles (the particles are taken to actually exist).   Thus, distinct pure states correspond to distinct ontic states and so the pilot wave model
is $\psi$-ontic.  The influence of the pilot wave model of de Broglie and Bohm on the study of quantum foundations has been significant.  Bell was motivated by the pilot wave model to ask whether any hidden variable model must
be nonlocal.  Valentini was motivated by the pilot wave model to ask whether any deterministic hidden variable model leads to signaling for non-equilibrium distributions of the hidden
variables \cite{valentini2002signal}.  Likewise, the question of whether any hidden variable model must be $\psi$-ontic is also motivated by the pilot wave model.  The pilot wave model
satisfies the restricted ontic indifference assumption (it is a matter of calculation as to whether it satisfies ontic indifference).

It is worth mentioning, as an aside, that the pilot wave model acts as a counterexample to the often encountered claim that quantum theory does not admit a hidden variable (or ontic) model.
It does.  The pilot wave model is one example.  Further, there exist versions of the pilot wave model capable of reproducing enough quantum (field) theory to account for the standard model
(see, for example, \cite{struyve2007minimalist}) so such models can account for all of present day quantum phenomena.

Other examples of $\psi$-ontic models are the many worlds interpretation \cite{everett1957relative}, collapse models in the standard ontology where the wavefunction is taken to be real
\cite{pearle1976reduction, ghirardi1986unified} (though the status of the flash ontology of Bell \cite{bell2004speakable} is not so clear), and the Beltrametti Bugajski model
\cite{beltrametti1995classical}.

There have been other takes on the question of whether the wavefunction is real.   Aharonov and Vaidman \cite{aharonov1993measurement} showed that it is possible to directly measure the wave
function with weak measurements using a protective measurement scheme given only a single copy of the system.  For this to work, however, it is necessary that the wavefunction is an
eigenstate of the acting Hamiltonian and so a measurement onto the energy basis would enable us to deduce the wavefunction. Nevertheless, the fact that the shape of the wavefunction can be
directly measured is quite remarkable.  Fuchs regards the quantum state as a degree of belief by taking a Bayesian interpretation of probabilities \cite{fuchs2010qbism}.  In particular, he
and collaborators have shown that there can be situations in which it is consistent for different agents to have different beliefs as to what the state of a given system is and yet for them
all to believe the system is in a pure state \cite{caves2002conditions}.  There is clearly a tension between that result and any proof that the wavefunction is a real thing which would be
interesting to explore further.  Indeed, one can think of a quantum state as being represented by a list of probabilities (corresponding to a complete set of fiducial measurements).  Then the only difference between two different pure quantum states is that they have different probabilities for these fiducial measurements.  If we adopt the point of view that probabilities are nothing more than degrees of belief then it is clearly possible for two agents to ascribe different pure states to the same situation.  It follows that there is a conflict between this view of probabilities and the three assumptions used in this paper.  Indeed, in the point of view developed by Fuchs \cite{fuchs2010qbism}, the first assumption (that there is a hidden variable $\lambda$ associated with a system) is not true.   Although this assumption was dubbed \lq\lq realism" we should admit that it is particular way of implementing the notion of realism.  We may have models do not deny realism in the broader sense of the word but which are inconsistent with this assumption.

If we restrict ourselves to the case of a spin half particle then there are two interesting models in the literature. Bell \cite{bell1964on} constructed a simple hidden variable model in
which distinct pure spin half states correspond to hidden variables represented by distinct unit vectors and the probability of seeing certain outcome is given by the overlap between this
vector and the vector representing the measurement direction.  This is a $\psi$-ontic model.  On the other hand, Kochen and Specker \cite{kochen1967problem} gave an example in which a pure
spin state corresponds to a distribution of unit vector hidden variables. This is a $\psi$-epistemic model.  Both these models are restricted to standard spin measurements on spin half
particles and do not work for Hilbert space dimension greater than 2.

An intermediate question we can ask is what the size of the ontic space must be to reproduce quantum theory.  The toy bit in the Spekkens toy theory has only four underlying ontic states.
However, although it behaves in a similar way to a qubit, it does not reproduce all predictions of quantum theory.  A preliminary result in this direction was provided by Galv\~ao and the
present author \cite{galvao2003substituting}.  It was shown that a single qubit can substitute for an infinite number of classical bits in a certain information processing task suggesting
that we need an infinite number of ontic states to model a single qubit.   Then it was proven by the present author \cite{hardy2004quantum} that the number of ontic states required to
reproduce quantum predictions for even a single qubit must, indeed, be infinite.  This proof depends critically on the fact that there are an infinite number of distinct pure quantum states
for a qubit (there exist an infinite number of states $|\psi\rangle = \alpha |0\rangle +\beta |1\rangle$ since $\alpha$ and $\beta$ vary continuously).

The proof in \cite{hardy2004quantum} only showed that there must be an infinite number of ontic states in models reproducing quantum theory.  As discussed in the introduction, a big step
forward was made by Montina \cite{montina2008exponential} when he showed, by making some very reasonable assumptions (most importantly he assumed that the model was Markovian), that the
hidden variables for a system with Hilbert space dimension $N$ must contain at least $2N-2$ real parameters - this being the same as the number of real parameters required to describe a
state $|\psi\rangle=\sum c_n|n\rangle$ after normalization and an over all phase has been set.  Further in \cite{montina2011state}, Montina showed that in models for which the ontic space is
specified by the minimum number of real parameters, the ontic state is isomorphic to a vector in an $N$ dimensional Hilbert space and, further, this vector evolves according to the
Schroedinger equation.  These results of Montina provide very strong evidence under rather weak assumptions that the quantum state is real. Even if the quantum state is not real, Montina's
results show that, in Markovian models, there is something at least as complicated that is real.

Montina's work falls just short of actually proving that quantum theory is $\psi$-ontic.  This was first proven by Pusey, Barrett, and Rudolph.  The result of PBR represents a significant
breakthrough in quantum foundations and has already been the subject of much study \cite{hall2011generalisations, colbeck2012system, miller2012alternative, drezet2012can,
schlosshauer2012recent} (see in particular Leifers illuminating discussion \cite{Leifer2011}).  The key assumption that PBR make to get their result is that the distribution over ontic
states for independently prepared pure quantum states should factorize.

It is worth remarking, in view of the general acclaim for the extraordinary PBR result that, in some respects, the results of Montina are stronger than those of PBR and, indeed, the present
paper.  By assuming less, Montina proves less.  However, his result is a bigger obstacle to finding worthwhile $\psi$-epistemic models - namely those in which the ontic space is simpler than
the pure quantum state space.  One of the biggest motivations to construct a $\psi$-epistemic model must be to achieve a simpler description of reality than is afforded by the quantum state.
It is possible that, by violating the  assumptions PBR and the present paper, we can build $\psi$-epistemic models. Further, by violating Montina's Markovian assumption, we may even be able
to build $\psi$-epistemic models that have fewer real variables than required to describe the quantum state.   Montina has shown that, for Hilbert space dimension $2$, it is possible to
build a non-Markovian model having only one (rather than two) real parameters in the ontic state. This theme has been continued by Montina in \cite{montina2011dynamics}. Lewis, Jennings,
Barrett, and Rudolph \cite{lewis2012quantum}, by violating the separability assumption of PBR, have shown how to build a $\psi$-epistemic model reproducing quantum theory for any dimension
of the Hilbert space (this model, however, requires as many real parameters as there are independent real parameters in the quantum state and it also violates the ontic indifference
assumption of the present paper).

The restricted ontic indifference assumption is a much stronger assumption than either Montina's Markovian assumption or PBR's separability assumption.  The real result of this paper is, then, that any $\psi$-epistemic model must violate restricted ontic indifference (and therefore also violate ontic indifference).  The fact that the Spekkens's toy model violates ontic indifference suggests that it is a step in the right direction.  As demonstrated by the pilot wave model for non-relativistic particles, $\psi$-ontic models need not violate restricted ontic indifference.  This property seems, then, to be a key difference between the two approaches (though, of course, $\psi$-ontic approaches do not have to satisfy ontic indifference).

One attitude one could take to the results of this paper and the work of Montina and PBR is that we should give up on trying to find interpretations of quantum theory in which the wave
function is not a real thing.   However, the benefits of finding an interpretation in which the wave function is not a real thing may far outweigh the problems caused by having to violate
the various assumptions used in the afore mentioned work.  In fact it is quite easy to violate the assumption of ontic indifference in hidden variable models.  Further, one can question how well motivated the PBR separability assumption (for independently prepared systems) is in view of the fact that, generically, we do not expect hidden variable interpretations of quantum theory to be separable when the systems are not independently prepared (see Spekkens discussion of the PBR result in \cite{spekkens2012why}).   Further, as just mentioned, Montina has already embarked on a project looking for models in which the number of real parameters
required for the ontic state is fewer than required to specify the quantum state (such models are necessarily $\psi$-epistemic).   Further, the work of Spekkens and collaborators
\cite{spekkens2007evidence, bartlett2011reconstruction} points to the advantages of a $\psi$-epistemic approach. An alternative, and even more radical, approach is that of Fuchs \cite{fuchs2010qbism} and collaborators. In this approach the quantum state is not a real thing since different agents can attribute different states for the same physical situation.

The area of realistic interpretations of quantum theory in which the quantum state is not a real thing remains relatively unexplored.  It
represents a very different take on the problem of interpretation to the old school approaches of de Broglie and Bohm (the pilot wave model), of Everett (the many worlds interpretation), and of Pearl, Ghirardi, Rimini, and Weber (the collapse models) from the last century.  The search for reasonable $\psi$-epistemic models (violating the assumptions implicit in the various no-go theorems) is likely to be come an increasingly exciting area of research in quantum foundations.

\section*{Acknowledgements}

I am especially grateful to Antony Valentini not only for numerous provocative discussions on the issue of reality in quantum theory over the years but also for providing limited funding
for the present project (\lq\lq As for the \$40 [you owe me], I'll make a deal with you. If you write another paper about hidden variables, I'll write off your debt ...").  I am also
grateful to Roberta Tevlin for inviting me to give a talk at the Ontario Association of Physics Teachers.  The idea for this paper arose in preparing that talk (in particular, in rehearsing
the argument that something goes both ways through an interferometer).  I am grateful to Robert Spekkens for numerous discussions and inspirational seminars on the subject of
$\psi$-epistemic models, to Jonathan Barrett for carefully explaining the assumptions going into the PBR proof (the first two assumptions here came from that discussion), and to Christopher Fuchs for discussions on Qbism.  Last, but certainly not least, I am very grateful to Matthew Leifer for pointing out that the toy model of Spekkens violate the ontic indifference assumption and, further, for pointing out that the interferometer version of this toy model due to Martin and Spekkens illustrates how a local violation of the restricted ontic indifference assumption is possible in a field ontology.  

Research at Perimeter Institute for Theoretical Physics is supported in part by the Government of Canada through NSERC and by the Province of Ontario through MRI.

\bibliography{psionticbib} \bibliographystyle{plain}

\end{document}